\documentclass[prc,aps,showpacs]{revtex4}
\usepackage{graphicx}
\begin{document}

\title{K$^+$-nucleon scattering and exotic S=+1 Baryons}

\author{R.~A.~Arndt}
\email{arndt@reo.ntelos.net}
\author{I.~I.~Strakovsky}
\email{igor@gwu.edu}
\author{R.~L.~Workman}
\email{rworkman@gwu.edu}
\affiliation{Center for Nuclear Studies, Department of
        Physics, \\
        The George Washington University, Washington,
        D.C. 20052, U.S.A.}

\begin{abstract}

The $K^+$-nucleon elastic scattering process has been 
reexamined in light of recent measurements which have
found a narrow exotic S=+1 resonance in their $KN$ 
invariant mass distributions.  We have analyzed the 
existing database in order to consider the effect of a 
narrow state on fits to $K^+$-nucleon observables.

\end{abstract}


\maketitle

An analysis of isoscalar and isovector contributions to $K^+$-nucleon
scattering, using $K^+p$ and $K^+d$ data, was published~\cite{vpi} by 
the VPI group in 1992.  The interest in this reaction was driven by 
a search for possible 
$Z^\ast$ resonances.  At the time of the VPI analysis, only isovector 
pole parameters had been reported in the Review of Particle Properties~
\cite{rpp} (RPP). Breit-Wigner parameters were available for two
isoscalar states. By including $K^+d$ elastic scattering and breakup 
data, an improved study of the isoscalar component was performed.  Both 
isovector and isoscalar poles were reported.  However, 
the inferred widths were of the order of 100~MeV, and could be 
``explained" as pseudo-resonances whose resonance-like behavior was 
induced by the opening of channels such as $K^+\Delta$ and $K^\ast 
N$. As a result, these states were removed from the RPP with the 
statement that ``the general prejudice against baryons not made of 
three quarks and the lack of any experimental activity'' precluded 
any foreseeable progress.

Recent measurements~\cite{spring,itep,jlab,elsa} from SPring-8, ITEP, 
Jefferson Lab, and ELSA have dramatically changed the status of 
$Z^\ast$ resonances (now denoted as $\Theta$ states). Unlike, the 
broad resonancelike structures seen in previous $K^+N$ analyses, 
these groups have reported very narrow peaks in their $KN$ mass 
distributions.  The reported masses have been consistent, clustered 
around 1540~MeV, with widths less than 25~MeV. These results are in 
remarkable agreement with a chiral soliton prediction of a $\Theta^+$ 
state at 1530 MeV, with a width less than 15 MeV~\cite{dpp}.

In order to understand how this state could have eluded previous 
studies of $K^+N$ scattering data, we have reanalyzed this reaction, 
focusing on the 1540~MeV region. While there are suggestions that 
this state should be seen in the P$_{01}$ partial wave, we have 
scanned for S-, P-, and D-wave ($S_{01}$, $S_{11}$, $P_{01}$,
$P_{03}$, $P_{11}$, $P_{13}$, $D_{03}$, $D_{05}$, $D_{13}$, and 
$D_{15}$) structures as well.  Width 
estimates have generally been given as upper limits. For this 
reason, we have allowed for much lower values.  As there have been 
essentially no new data for this reaction since our last published 
fit, we have used the 1992 result as a starting point.  The fitting 
strategy and database are fully documented in Ref.~\cite{vpi}. Below, 
we give only our results from a search for narrow structures.

Narrow resonances were added to the VPI analysis using a product 
S-matrix approach. The added state was assumed to be an elastic
$KN$ resonance with an energy-independent 
width.  An additional background S-matrix was also included, $S_B \; 
= \; (1+iK)/(1-iK)$, with an underlying K-matrix proportional to 
phase space.  Data were then analyzed with resonances covering a 
grid of mass and width values, in order to see which values were
favored~\cite{azimov} in a particular partial wave.  From an initial 
scan with masses in the range between 1520 and 1560~MeV, with widths 
from 1 to 10~MeV, one feature was immediately obvious.  The addition 
of states with widths above 5~MeV resulted in an enormous increase 
in $\chi^2$, often more than doubling the $\chi^2$/data found in a 
fit from threshold to T$_{\rm Lab}$ = 1.1~GeV  ($\chi^2$/data = 
3198/1746) for isoscalar waves, and threshold to 
T$_{\rm Lab}$ = 2.65~GeV ($\chi^2$/data = 4872/3663) for isovector waves.
The effect on $\chi^2$ is illustrated in Fig.~
\ref{fig:g1}a for the $P_{01}$ partial wave.  Similar results were 
found for other partial waves. This is in agreement with a claim by 
Nussinov~\cite{nussinov}, based on a lack of pronounced structure 
in the $K^+d$ total cross section.  In Ref.~\cite{nussinov}, the 
conservative estimate $\Gamma_{\Theta^+}$ $<$ 6~MeV was given.  A 
limit of $\Gamma_{\Theta^+}$ $<$ 9~MeV was given in the
ITEP bubble chamber measurements of Ref.~\cite{itep}.

We found no $\chi^2$ improvement associated with the addition of 
resonances in the S-, P- and D-waves, unless the inserted structures 
had widths of order 1~MeV or less.  However, for very narrow widths, 
a lack of data at energies corresponding to the $\Theta^+$ was a 
limiting factor. [In our search for narrow $N'$ states in the $\pi N$ 
elastic scattering reaction, we found that it was generally possible 
to insert keV-width states, without serious effect, if they filled 
gaps in the database~\cite{azimov}.]  In Fig.~\ref{fig:g1}b, we 
again plot the change in $\chi^2$ associated with resonances (in the 
$P_{01}$ wave) having widths below 1~MeV. The location of available data is 
also indicated.  Unfortunately, the $\Theta^+$ mass, near 
1545~MeV, falls in a data gap.  As a result, we have little 
sensitivity to the $\Theta^+$, if its width is much below an MeV. In 
Fig.~\ref{fig:g2}, we show similar mappings for the $S_{01}$ and
$P_{03}$ partial waves, also for the case of sub-MeV widths. The
$P_{01}$ plots have the most suggestive dips near the $\Theta^+$ mass, 
but with such sparse data coverage, this cannot be considered as
definitive proof.  Plots of isovector waves for the widths displayed 
in Figs.~\ref{fig:g1},~\ref{fig:g2} show no regions of negative $\Delta\chi^2$.

In summary, based upon a reanalysis of the existing $K^+N$ database, 
we find that $\Theta^+$ widths beyond the few-MeV level are excluded,
confirming a claim based on total cross section data~\cite{nussinov}. 
The existence of a $\Theta^+$ in the $P_{01}$ state, with a width of 
an MeV or less, is possible.  We see no evidence
for a $\Theta^{++}$ in the existing $K^+p$ scattering data. As the $\Theta^+$
has been seen in both $K^+ n$ and $K^0 p$ mass distributions, our assumption
of an elastic process in $KN$ scattering seems reasonable. 
However, it appears that a more definitive conclusion will have to await further
measurements.

\acknowledgments

The authors thank Ya.~Azimov, A.~Dolgolenko, and W.~Roberts for many helpful 
discussions.  This work was supported in part by the U.~S. Department 
of Energy grant DE--FG02--99ER41110, and by Jefferson Lab and the 
Southeastern Universities Research Association under DOE contract 
DE--AC05--84ER40150.

\end{document}